Submitted to *Communications of the ACM*

# Comment on: "Was Ada Lovelace Actually the First Programmer?"

Thomas J. Misa

November 16, 2022

Additional resources are available to *Communications of the ACM* readers intrigued by Herbert Bruderer's recent, "Was Ada Lovelace Actually the First Programmer?" (online) which is a pert condensing of his recent book's treatment on pages 1002-3: see *Milestones in Analog and Digital Computing* (Springer, 3rd edition, 2020) (online). With *Ada's Legacy: Cultures of Computing from the Victorian to the Digital Age*, edited by Robin Hammerman and Andrew L. Russell, ACM Books published Lovelace's entire 1843 article (translation and explanatory notes). There readers can find "Sketch of the Analytical Engine invented by Charles Babbage," *Scientific Memoirs (1843)* (online), and on page 105 an exact facsimile of Lovelace's "Diagram for the computation by the Engine of the Numbers of Bernoulli" which clearly shows two nested loops in the tabular computation [below]. A link on my chapter's page 20 points to a high-resolution image of this diagram (online and archived).

Diagram for the computation by the Engine of the Numbers of Bernoulli. See Note G. (page 722 et seq.)

| Number of Operation. | Nature of Operation. | Variables acted upon. | Variables receiving results. | Indication of change in the value on any Variable. | Statement of Results. | Data. $^1V_1$ 0 0 0 1 [1] | $^1V_2$ 0 0 0 2 [2] | $^1V_3$ 0 0 0 4 [n] | Working Variables. $^0V_4$ 0 0 0 0 | $^0V_5$ 0 0 0 0 | $^0V_6$ 0 0 0 0 | $^0V_7$ 0 0 0 0 | $^0V_8$ 0 0 0 0 | $^0V_9$ 0 0 0 0 | $^0V_{10}$ 0 0 0 0 | $^0V_{11}$ 0 0 0 0 | $^0V_{12}$ 0 0 0 0 | $^0V_{13}$ 0 0 0 0 | Result Variables. $^1V_{21}$ $B_1$ in a decimal fraction. [$B_1$] | $^1V_{22}$ $B_3$ in a decimal fraction. [$B_3$] | $^1V_{23}$ $B_5$ in a decimal fraction. [$B_5$] | $^0V_{24}$ 0 0 0 0 [$B_7$] |
|---|---|---|---|---|---|---|---|---|---|---|---|---|---|---|---|---|---|---|---|---|---|---|
| 1 | × | $^1V_2 \times {}^1V_3$ | $^1V_4, {}^1V_5, {}^1V_6$ | $\{^1V_2 = {}^1V_2;\ ^1V_3 = {}^1V_3\}$ | $= 2n$ | ... | 2 | $n$ | $2n$ | $2n$ | $2n$ | | | | | | | | | | | |
| 2 | − | $^1V_4 - {}^1V_1$ | $^2V_4$ | $\{^1V_4 = {}^2V_4;\ ^1V_1 = {}^1V_1\}$ | $= 2n - 1$ | 1 | ... | ... | $2n-1$ | | | | | | | | | | | | | |
| 3 | + | $^1V_5 + {}^1V_1$ | $^2V_5$ | $\{^1V_5 = {}^2V_5;\ ^1V_1 = {}^1V_1\}$ | $= 2n + 1$ | 1 | ... | ... | ... | $2n+1$ | | | | | | | | | | | | |
| 4 | ÷ | $^2V_5 \div {}^2V_4$ | $^1V_{11}$ | $\{^2V_5 = {}^0V_5;\ ^2V_4 = {}^0V_4\}$ | $= \frac{2n-1}{2n+1}$ | ... | ... | ... | 0 | 0 | ... | ... | ... | ... | ... | $\frac{2n-1}{2n+1}$ | | | | | | |
| 5 | ÷ | $^1V_{11} \div {}^1V_2$ | $^2V_{11}$ | $\{^1V_{11} = {}^2V_{11};\ ^1V_2 = {}^1V_2\}$ | $= \frac{1}{2} \cdot \frac{2n-1}{2n+1}$ | ... | 2 | ... | ... | ... | ... | ... | ... | ... | ... | $\frac{1}{2} \cdot \frac{2n-1}{2n+1}$ | | | | | | |
| 6 | − | $^0V_{13} - {}^2V_{11}$ | $^1V_{13}$ | $\{^2V_{11} = {}^0V_{11};\ ^0V_{13} = {}^1V_{13}\}$ | $= -\frac{1}{2} \cdot \frac{2n-1}{2n+1} = A_0$ | ... | ... | ... | ... | ... | ... | ... | ... | ... | ... | 0 | ........ | $-\frac{1}{2} \cdot \frac{2n-1}{2n+1} = A_0$ | | | | |
| 7 | − | $^1V_3 - {}^1V_1$ | $^1V_{10}$ | $\{^1V_3 = {}^1V_3;\ ^1V_1 = {}^1V_1\}$ | $= n - 1\ (= 3)$ | 1 | ... | $n$ | ... | ... | ... | ... | ... | ... | $n-1$ | | | | | | | |
| 8 | + | $^1V_2 + {}^0V_7$ | $^1V_7$ | $\{^1V_2 = {}^1V_2;\ ^0V_7 = {}^1V_7\}$ | $= 2 + 0 = 2$ | ... | 2 | ... | ... | ... | ... | 2 | | | | | | | | | | |
| 9 | ÷ | $^1V_6 \div {}^1V_7$ | $^3V_{11}$ | $\{^1V_6 = {}^1V_6;\ ^0V_{11} = {}^3V_{11}\}$ | $= \frac{2n}{2} = A_1$ | ... | ... | ... | ... | ... | $2n$ | 2 | ... | ... | ... | $\frac{2n}{2} = A_1$ | | | | | | |
| 10 | × | $^1V_{21} \times {}^3V_{11}$ | $^1V_{12}$ | $\{^1V_{21} = {}^1V_{21};\ ^3V_{11} = {}^3V_{11}\}$ | $= B_1 \cdot \frac{2n}{2} = B_1 A_1$ | ... | ... | ... | ... | ... | ... | ... | ... | ... | ... | $\frac{2n}{2} = A_1$ | $B_1 \cdot \frac{2n}{2} = B_1 A_1$ | ........ | $B_1$ | | | |
| 11 | + | $^1V_{12} + {}^1V_{13}$ | $^2V_{13}$ | $\{^1V_{12} = {}^0V_{12};\ ^1V_{13} = {}^2V_{13}\}$ | $= -\frac{1}{2} \cdot \frac{2n-1}{2n+1} + B_1 \cdot \frac{2n}{2}$ | ... | ... | ... | ... | ... | ... | ... | ... | ... | ... | ........ | 0 | $\left\{-\frac{1}{2} \cdot \frac{2n-1}{2n+1} + B_1 \cdot \frac{2n}{2}\right\}$ | | | | |
| 12 | − | $^1V_{10} - {}^1V_1$ | $^2V_{10}$ | $\{^1V_{10} = {}^2V_{10};\ ^1V_1 = {}^1V_1\}$ | $= n - 2\ (= 2)$ | 1 | ... | ... | ... | ... | ... | ... | ... | ... | $n-2$ | | | | | | | |
| 13 | − | $^1V_6 - {}^1V_1$ | $^2V_6$ | $\{^1V_6 = {}^2V_6;\ ^1V_1 = {}^1V_1\}$ | $= 2n - 1$ | 1 | ... | ... | ... | ... | $2n-1$ | | | | | | | | | | | |
| 14 | + | $^1V_1 + {}^1V_7$ | $^2V_7$ | $\{^1V_1 = {}^1V_1;\ ^1V_7 = {}^2V_7\}$ | $= 2 + 1 = 3$ | 1 | ... | ... | ... | ... | ... | 3 | | | | | | | | | | |
| 15 | ÷ | $^2V_6 \div {}^2V_7$ | $^1V_8$ | $\{^2V_6 = {}^2V_6;\ ^2V_7 = {}^2V_7\}$ | $= \frac{2n-1}{3}$ | ... | ... | ... | ... | ... | $2n-1$ | 3 | $\frac{2n-1}{3}$ | | | | | | | | | |
| 16 | × | $^1V_8 \times {}^3V_{11}$ | $^4V_{11}$ | $\{^1V_8 = {}^0V_8;\ ^3V_{11} = {}^4V_{11}\}$ | $= \frac{2n}{2} \cdot \frac{2n-1}{3}$ | ... | ... | ... | ... | ... | ... | ... | 0 | ... | ... | $\frac{2n}{2} \cdot \frac{2n-1}{3}$ | | | | | | |
| 17 | − | $^2V_6 - {}^1V_1$ | $^3V_6$ | $\{^2V_6 = {}^3V_6;\ ^1V_1 = {}^1V_1\}$ | $= 2n - 2$ | 1 | ... | ... | ... | ... | $2n-2$ | | | | | | | | | | | |
| 18 | + | $^1V_1 + {}^2V_7$ | $^3V_7$ | $\{^2V_7 = {}^3V_7;\ ^1V_1 = {}^1V_1\}$ | $= 3 + 1 = 4$ | 1 | ... | ... | ... | ... | ... | 4 | | | | | | | | | | |
| 19 | ÷ | $^3V_6 \div {}^3V_7$ | $^1V_9$ | $\{^3V_6 = {}^3V_6;\ ^3V_7 = {}^3V_7\}$ | $= \frac{2n-2}{4}$ | ... | ... | ... | ... | ... | $2n-2$ | 4 | ... | $\frac{2n-2}{4}$ | ... | $\left\{\frac{2n}{2} \cdot \frac{2n-1}{3} \cdot \frac{2n-2}{3} = A_3\right\}$ | | | | | | |
| 20 | × | $^1V_9 \times {}^4V_{11}$ | $^5V_{11}$ | $\{^1V_9 = {}^0V_9;\ ^4V_{11} = {}^5V_{11}\}$ | $= \frac{2n}{2} \cdot \frac{2n-1}{3} \cdot \frac{2n-2}{4} = A_3$ | ... | ... | ... | ... | ... | ... | ... | ... | 0 | | | | | | | | |
| 21 | × | $^1V_{22} \times {}^5V_{11}$ | $^0V_{12}$ | $\{^1V_{22} = {}^1V_{22};\ ^0V_{12} = {}^2V_{12}\}$ | $= B_3 \cdot \frac{2n}{2} \cdot \frac{2n-1}{3} \cdot \frac{2n-2}{3} = B_3 A_3$ | ... | ... | ... | ... | ... | ... | ... | ... | ... | ... | 0 | $B_3 A_3$ | ........ | ...... | $B_3$ | | |
| 22 | + | $^2V_{12} + {}^2V_{13}$ | $^3V_{13}$ | $\{^2V_{12} = {}^0V_{12};\ ^2V_{13} = {}^3V_{13}\}$ | $= A_0 + B_1 A_1 + B_3 A_3$ | ... | ... | ... | ... | ... | ... | ... | ... | ... | ... | ........ | 0 | $\{A_0 + B_1 A_1 + B_3 A_3\}$ | | | | |
| 23 | − | $^2V_{10} - {}^1V_1$ | $^3V_{10}$ | $\{^2V_{10} = {}^3V_{10};\ ^1V_1 = {}^1V_1\}$ | $= n - 3\ (= 1)$ | 1 | ... | ... | ... | ... | ... | ... | ... | ... | $n-3$ | | | | | | | |
| Here follows a repetition of Operations thirteen to twenty-three. | | | | | | | | | | | | | | | | | | | | | | |
| 24 | + | $^4V_{13} + {}^0V_{24}$ | $^1V_{24}$ | $\{^4V_{13} = {}^0V_{13};\ ^0V_{24} = {}^1V_{24}\}$ | $= B_7$ | ... | ... | ... | ... | ... | ... | ... | ... | ... | ... | ........ | ........ | ........ | ...... | ...... | ...... | $B_7$ |
| 25 | + | $^1V_1 + {}^1V_3$ | $^1V_3$ | $\{^1V_1 = {}^1V_1;\ ^1V_3 = {}^1V_3;\ ^5V_6 = {}^0V_6;\ ^5V_7 = {}^0V_7\}$ | $= n + 1 = 4 + 1 = 5$; by a Variable-card.; by a Variable-card. | 1 | ... | $n+1$ | ... | ... | 0 | 0 | | | | | | | | | | |

Other resources stem from the efforts of Ursula Martin and colleagues at Oxford who organized a December 2015 Lovelace celebration. These include *Ada Lovelace: The Making of a Computer Scientist* by Christopher Hollings, Ursula Martin, Adrian Rice (Bodleian Libraries, 2018 see here). A selection of Ada Lovelace correspondence is also online.

Historical analysis of any complex topic involves matters of professional judgment rather than neat proof or polling. Bruderer cites two Babbage authorities, Doron Swade and Allan Bromley, to support his assertion that Lovelace is "celebrated unjustly" as the first computer programmer. Allan Bromley spent many years celebrating the originality and achievements of Charles Babbage, and for years dismissed Ada Lovelace owing to her (presumed) lack of mathematical sophistication. Bromley cast a long and influential shadow. But recent research on the mathematically sophisticated correspondence between Lovelace and the celebrated Augustus De Morgan — see Hollings et al. "The Lovelace-De Morgan Mathematical Correspondence: A Critical Re-Appraisal," *Historia Mathematica* 2017 (online) and "The Early Mathematical Education of Ada Lovelace" *BSHM Bulletin* 2017 (online) — surely must revise that assertion. Since the concept "to program" a computer dates from decades later, I agree with Bruderer that labeling Lovelace as the "first programmer" is unhelpful and anachronistic. I believe the archival evidence pointing to Ada Lovelace's authorship of the step-by-step—literally, algorithmic—means to compute the series of Bernoulli numbers correctly describes her achievement.

Thomas J. Misa
tmisa@umn.edu